\documentclass[sigconf]{acmart}
\usepackage{todonotes}
\usepackage{graphicx}

\copyrightyear{2020}
\acmYear{2020}
\setcopyright{rightsretained}
\acmConference[WSDM '20]{The Thirteenth ACM International Conference on Web Search and Data Mining}{February 3--7, 2020}{Houston, TX, USA}
\acmBooktitle{The Thirteenth ACM International Conference on Web Search and Data Mining (WSDM '20), February 3--7, 2020, Houston, TX, USA}

\begin{document}
\fancyhead{}


\title{User Intent Inference for Web Search and Conversational Agents}

\author{Ali Ahmadvand }
\affiliation{ \institution{Computer Science Department \\Emory University, Atlanta, GA, USA}}
\email{ali.ahmadvand@emory.edu}

\begin{abstract}
User intent understanding is a crucial step in designing both conversational agents and search engines. Detecting or inferring user intent is challenging, since the user utterances or queries can be short, ambiguous, and contextually dependent. To address these research challenges, my thesis work focuses on: 1) Utterance topic and intent classification for conversational agents 2) Query intent mining and classification for Web search engines, focusing on the e-commerce domain. To address the first topic, I proposed novel models to incorporate entity information and conversation-context clues to predict both topic and intent of the user's utterances. For the second research topic, I plan to extend the existing state of the art methods in Web search intent prediction to the e-commerce domain, via: 1) Developing a joint learning model to predict search queries' intents and the product categories associated with them, 2) Discovering new hidden users' intents. All the models will be evaluated on the real queries available from a major e-commerce site search engine. The results from these studies can be leveraged to improve performance of various tasks such as natural language understanding, query scoping, query suggestion, and ranking, resulting in an enriched user experience.

\end{abstract}

\keywords{User Intent Understanding; User intent in Conversational Agents; User Intent in E-Commerce Search }

\maketitle

\vspace{-0.2cm}
\section{Background and Motivation}

Understanding the user's intents for general Web search engines have been well studied in the last decades\cite{Broder:2002, Jian:2009}. However, for E-Com domain, user intent understanding is in its initial stages \cite{Parikshit:2018}. In recent years, with the drastic increase of online shopping and E-Com sales growth, there is a high demand for designing effective user intent understanding models to optimize the search results, which can directly impact different business metrics such as gross demand. 
Furthermore, developing an effective conversational search that can interact with users in a conversation-like manner has been a long-term goal in designing a search engine. Recently, with the new developments in speech recognition technology \cite{Wayne:2018}, companies like Amazon goes further, in that they invest to develop their own socialbots that can keep a coherent conversation in an open-domain and human-like manner with their customers \cite{rf:emerson, rf:iris}. As a result, developing an accurate and efficient Natural Language Understanding (NLU) unit for these socialbots has been in the spotlight in recent years ~\citep{Guo:ADAN}.


The new problems for the user intent understanding raised by these new applications open opportunities for researchers to develop new ideas in these fields. 
To keep bridging these gaps, my Ph.D. dissertation focuses on user intent understanding for either conversational agents and E-Com search engines, which has been done with collaboration with Dr. Eugene Agichtein. My dissertation consists of two phases of research development. The first part of my dissertation focused on developing multiple machine learning models to develop a NLU component for conversational agents. 
To this end, my main research objectives were as follows:

\begin{enumerate}
    \item Incorporating contextual information into intent classification in an open-domain conversational agent.
    \item Developing a method to utilize the knowledge available from human conversations for a human-machine conversation.
    \item Implementing an entity-aware topic classification model for an open-domain conversational agent.
    \item Developing an online satisfaction prediction mechanism in an open-domain conversational agent.
\end{enumerate}

In the second phase of my dissertation, I plan to address specific issues related to user intent mining and classification for E-Com search engines. User intent understanding for search engines is similar to conversational bots. However, for the search engines, there are much more data and user behavior information such as click rates available than conversational bots. My main research questions leading my work are as follows:

\begin{enumerate}
    \item How efficient is it to develop a joint learning model to simultaneously learn the search query's product category and user intents?
    \item How can we discover users' hidden intents encoded in E-com search logs?
\end{enumerate}

\vspace{-0.2cm}
\section{Overview of Research Directions}
In this section, my research direction and different projects relevant to it are described. The first section is an overview of the Amazon Alexa Prize competitions\footnote{\textit{\url{https://developer.amazon.com/alexaprize}}} and my contributions within them. In the second section, my primary research on user intent understanding for E-Com search engines is discussed.


\vspace{-0.3cm}
\subsection{Amazon Alexa Prize 2017 and 2018}
For Amazon Alexa Prize 2017 \cite{rf:emerson}, our team developed one of the first socialbots named Emersonbot that can converse with general users in an open-domain manner. Our approach to solve such a challenge was considering a conversation as a search engine, while the input consists of human utterances. 

For Alexa Prize 2018 \cite{rf:iris}, much like Alexa Prize 2017, we developed an open domain socialbot named IrisBot that can converse in multiple domains with real users. We improved the capability of IrisBot with respect to EmersonBot in several aspects and I published several papers documenting this work. First, for user intent classification, I proposed a novel method, CDAC (Contextual Dialogue Act Classifier), a simple yet effective deep learning approach for contextual dialogue act (broad intent) classification. Specifically, we used transfer learning to adapt models trained on human-human conversations to predict dialogue acts in human-machine dialogues \cite{rf:CDAC}. Then, for topic classification, we introduced a Concurrent Conversational Entity-aware Topic classifier (ConCET), which incorporates entity-type information together with the utterance content features. Specifically, ConCET utilizes entity information to enrich the utterance representation, combining character, word, and entity-type embeddings into a single representation \cite{rf:concet}. Finally, we proposed a Conversational Satisfaction prediction model specifically designed for open-domain conversational agents, called ConvSAT. To operate robustly across domains, ConvSAT aggregates multiple representations of the conversation, namely the conversation history, utterance and response content, and system- and user-oriented behavioral signals \cite{rf:convsat}.

\vspace{-0.3cm}
\subsection{E-Commerce Search Engines}
This part of my research was done during my internship in The Home Depot (THD)\footnote{\textit{\url{https://www.homedepot.com/}}}. THD receives billions of search queries every year, and collects terabytes of data logs from user experience during interaction with their website. My research focused on enhancing their high-level user intent understanding module. To this end, we proposed a hierarchical architecture for the user intent classification, where in the first layer, the intent of the users in purchasing a product or seeking information (product vs informational) is discovered. Then, in the next layer, if the user intent is purchasing, another intent classifier was applied to determine whether the query is either broad or specific. Otherwise, if the intent of the user is information seeking, another classifier determines the type of information the user is looking for. As a result, the search engine can guide a user to an appropriate web page to handle the user's request. 


\vspace{-0.3cm}
\section{Future Research}
\label{sec:future}
For the remaining part of my dissertation, I plan to extend my research to answer a couple of questions raised while working with a large E-Com search engine like that of The Home Depot.  

\vspace{-0.2cm}
\subsubsection*{\bf Boosting Search Performance Using Joint Learning. } Joint and multitask learning are crucial for an E-Com search engine in both engineering and performance perspectives. It is important from the engineering perspective because only one classifier is deployed instead of multiples, this contributes to reducing overhead and enhances maintenance of the system. It is also effective for implementation of relevant intent classification tasks due to the transferring of the inductive bias between two tasks. In the E-Com domain, accurate classification of queries will help with identifying the right product groupings from which relevant products can be retrieved. Additionally, search queries in this domain tend to be short, ambiguous, and the vocabulary tends to change as the catalog evolves. For product search, it might be necessary to identify intents associated with a query across various granularities such as category intent, accessory intent, vertical intent, etc. These intent classification tasks are related, knowledge from one task might help improve the performance of other tasks. To this end, I plan to design a joint learning model of high-level intent mapping (product vs. informational), and product category mapping as well as informational type classification to improve the performance of these three individual intent classification tasks.

\vspace{-0.2cm}
\subsubsection*{\bf Hidden User Intent Mining and Discovery. } The proposed hierarchical intent architecture
is effective, however, there might be new users' intents that are hidden. Providing a customized and fine-grained intent hierarchy model will assist an E-Com search engine to implement more efficiently intent classifiers, and consequently improve the user intent understanding. Unfortunately, existing research for E-Com setting in this topic is very limited. In an effort to advance this research field, researchers at Walmart introduced a clustering model \cite{Parikshit:2018} to extract the similarities between the users' behavior information while exploring the website. They suggested five main clusters only based on the behavioral data for their commercial website. This model solely relies on user behavior data while they lose the information incorporated in the query semantics. In contrast, in designing an NLU unit for spoken dialogue systems like Amazon Alexa, query semantics are the preliminary and dominant source of information. To this end, in this project, inspired by an NLU unit in conversational bots, I plan to develop a new model to incorporate the query semantics into the behavior data to find new hidden users' intents.

\vspace{-0.2cm}
\subsubsection*{\bf \textit{Acknowledgements}} We gratefully acknowledge the financial and computing support from the Amazon Alexa team and The Home Depot data science team during my Ph.D. dissertation.

\vspace{-0.3cm}
\bibliographystyle{ACM-Reference-Format}
\bibliography{References}

\end{document}